\documentclass[9pt,conference]{IEEEtran}
\IEEEoverridecommandlockouts

\usepackage[english]{babel}

\usepackage{atbegshi}
\usepackage{amsfonts}
\usepackage{algorithmic}
\usepackage{booktabs}
\usepackage{verbatim}
\usepackage{subcaption}

\usepackage{listings} 
\usepackage[frozencache=true,cachedir=minted-cache]{minted}
\usepackage{xcolor} 
\usepackage{multicol} 
\usepackage{float}

\newfloat{codeblock}{htbp}{loc}[section]
\floatname{codeblock}{Codeblock}

\usepackage[]{minted} 
\usepackage{newfloat}
\DeclareFloatingEnvironment[
    fileext=oluc,
    name=Listing
]{plisting}
\makeatletter
\newcommand\notsotiny{\@setfontsize\notsotiny\@vipt\@viipt}
\makeatother

\usepackage{multicol} 
\usepackage[para]{threeparttable}

\usepackage{color}
\usepackage{cellspace}
\usepackage{svg} 
\usepackage{placeins} 

\usepackage{siunitx} 
\usepackage{xfrac} 
\usepackage{xspace} 
\usepackage{footnote}
\usepackage{textgreek}
\usepackage{multirow}
\usepackage[nolist]{acronym} 
\usepackage[inline]{enumitem}
\usepackage{cleveref}

\definecolor{lightgray}{gray}{0.98}
\definecolor{cpu}{HTML}{434343}
\definecolor{carus}{HTML}{007480}
\definecolor{caesar}{HTML}{6d1a36}
\newcommand{\arcane}{ARCANE\xspace}
\newcommand{\at}{\ac{at}\xspace}

\newcommand{\caos}{\ac{cos}\xspace}
\newcommand{\cpu}{\ac{cpu}\xspace}
\newcommand{\ct}{\ac{ct}\xspace}
\newcommand{\dma}{\ac{dma}\xspace}

\newcommand{\imc}{\ac{imc}\xspace}

\newcommand{\isa}{\ac{isa}\xspace}

\newcommand{\llc}{\ac{llc}\xspace}

\newcommand{\mcu}{\ac{mcu}\xspace}
\acrodefplural{mcu}[MCUs]{Microcontroller Units}

\newcommand{\nmc}{\ac{nmc}\xspace}
\newcommand{\ooo}{\ac{ooo}\xspace}

\newcommand{\riscv}{RISC-V\xspace}

\newcommand{\sram}{\ac{sram}\xspace}

\acrodefplural{ann}[ANNs]{Artificial Neural Networks}

\newcommand{\ecpu}{\ac{ecpu}\xspace}
\newcommand{\vpu}{\ac{vpu}\xspace}
\newcommand{\vpus}{\acp{vpu}\xspace}
\newcommand{\xif}{\ac{xif}\xspace}
\newcommand{\emem}{\ac{emem}\xspace}
\newcommand{\xisa}{\texttt{xmnmc}\xspace}
\newcommand{\gops}{GOPS\xspace}

\newcommand{\tnotemid}[1]{\textsuperscript{\TPTtagStyle{#1}}} 

\newenvironment{acks}{%
  \section*{\textsc{Acknowledgements}}
}{}

\begin{document}


\title{ARCANE: Adaptive RISC-V Cache Architecture for Near-memory Extensions}

\author{%
\IEEEauthorblockN{Vincenzo Petrolo\IEEEauthorrefmark{1}}
\and
\IEEEauthorblockN{Flavia Guella\IEEEauthorrefmark{1}}
\and
\IEEEauthorblockN{Michele Caon\IEEEauthorrefmark{1}}
\and
\IEEEauthorblockN{Pasquale Davide Schiavone\IEEEauthorrefmark{2}}
\and
\IEEEauthorblockN{Guido Masera\IEEEauthorrefmark{1}}
\and
\IEEEauthorblockN{Maurizio Martina\IEEEauthorrefmark{1}}
\and
\IEEEauthorblockA{\IEEEauthorrefmark{1}%
\textit{VLSI Lab} \\
\textit{Politecnico di Torino, Italy}\\
\{vincenzo.petrolo,flavia.guella,michele.caon,guido.masera,maurizio.martina\}@polito.it}
\and
\IEEEauthorblockA{\IEEEauthorrefmark{2}%
\textit{Embedded Systems Laboratory} \\
\textit{EPFL, Switzerland}\\
davide.schiavone@epfl.ch}
\thanks{\textit{V. Petrolo and F. Guella contributed equally to this work.}}
}

\maketitle

\begin{abstract}
    Modern data-driven applications expose limitations of von Neumann architectures—extensive data movement, low throughput, and poor energy efficiency. Accelerators improve performance but lack flexibility and require data transfers. Existing compute in- and near-memory solutions mitigate these issues but face usability challenges due to data placement constraints.
    We propose a novel cache architecture that doubles as a tightly-coupled compute-near-memory coprocessor. Our \riscv cache controller executes custom instructions from the host CPU using vector operations dispatched to near-memory vector processing units within the cache memory subsystem. This architecture abstracts memory synchronization and data mapping from application software while offering software-based \acl*{isa} extensibility.
    Our implementation shows \SI{30}{\times} to \SI{84}{\times} performance improvement when operating on 8-bit data over the same system with a traditional cache when executing a worst-case 32-bit CNN workload, with only \SI{41.3}{\percent} area overhead.
\end{abstract}


\maketitle


\section{Introduction}


Modern computing systems are increasingly demanding higher performance and energy efficiency due to data-intensive workloads. Traditional von Neumann architectures struggle with scaling due to the so-called ``memory wall'' \cite{memory_wall} bottleneck, impacting performance and energy efficiency. To overcome these limitations, alternative computing paradigms like \ac{cim}  have been explored. \imc integrates processing within the memory array, reducing data movement and enhancing performance and energy efficiency. However, it faces challenges with complex arithmetic operations and memory density. \nmc, using conventional digital flows and commercial memory macros, offers better scalability and reliability compared to \imc.
\arcane, the architecture proposed in this work, builds upon the \nmc paradigm, aiming to bridge the gap between the flexibility of \cpu-based computing and the energy and performance benefits of \ac{cim}. It leverages a tightly coupled instruction offloading mechanism based on the OpenHW Group \xif coprocessor interface \cite{cvxif} to offload complex instructions to a \riscv cache controller, which implements them as vector kernels exploiting the \ac{cim}-oriented custom \isa extension from \cite{carus}. In other words, the \arcane cache system doubles as a tightly coupled coprocessor that abstracts complex in-cache computing operations inside a software-defined \isa that wraps the underlying custom vector \isa used to program efficient near-memory \acp{vpu} constituting the cache memory space.
The major contributions of the \arcane cache architecture are:
\begin{itemize}
    \item Efficient operation offloading: \arcane leverages a \cpu-based controller and a software-defined \isa to abstract complex in-cache operations, that are exposed to the host \cpu as complex instructions, offloaded through a coprocessor interface.
    \item Seamless integration with existing systems: \arcane is designed as a drop-in replacement for the conventional on-chip \ac{llc} of a host \ac{mcu}, minimizing the integration effort and ensuring compatibility with existing platforms.
    \item High-performance computing: \arcane's \acp{vpu} operate directly on data residing in the \ac{llc}, achieving \SI{30}{\times} to \SI{80}{\times} higher throughput compared to a \cpu-based approach when executing a 3-channel convolutional layer on 8-bit data.
\end{itemize}
The rest of this paper is organized as follows: \cref{sec:related} introduces relevant works from the literature; \cref{sec:arch} describes and motivates \arcane's architecture; \cref{sec:sw} details the customizable \isa and the runtime running in the cache controller; \cref{sec:results} reports and discusses implementation and performance figures; finally \cref{sec:conclusion} concludes the paper.

\section{Related Works}
\label{sec:related}


\ac{cim} approaches embedding processing elements within cache hierarchies have the inherent benefit of eliminating data transfers between system memory and dedicated compute units.
\sram-based \ac{imc} solutions \cite{blade,compute_caches,maic,dual_cache} repurpose the \sram blocks inside the cache as \ac{simd} accelerators exploiting bit-line computing and achieving high area and energy efficiency at the cost of reduced memory density and limited operational flexibility compared to conventional, off-the-shelf \acp{sram} \cite{blade_gem5x}.

In contrast, \nmc solutions place arithmetic units outside memory subarrays, leveraging denser, commercial \sram arrays and conventional digital implementation flow for greater portability.
These systems achieve higher throughput and energy efficiency by processing data close to memory, avoiding costly transfers via system interconnect.
Typical \nmc architectures rely on data-parallel execution units that operate on vectors \cite{intel_cache} or matrices \cite{nori2021reduct} to accelerate \ac{mac} operations in neural network inference.


Software integration challenges limit \ac{cim} commercial diffusion \cite{arxiv_landscape_PIM}.
\ac{imc} systems encode instructions for the memory as conventional bus transactions \cite{blade}, thus requiring the application software or a dedicated compiler to generate the necessary commands.
A more streamlined software integration can be achieved by connecting the \ac{cim} device to the host \cpu through a dedicated instruction offloading interface similar to a coprocessor \cite{intel_cache}. 
This approach facilitates the insertion of compute instructions into the generated application code, and guarantees synchronization with other in-flight instructions, implicitly handled by the \cpu.

The architecture proposed in this work builds upon the \nmc integration paradigm proposed with NM-Carus in \cite{carus} and the instruction-level offloading mechanism used in \cite{intel_cache} to relieve the application software from explicitly handling memory management and synchronization with the cache.

\section{Architecture}
\label{sec:arch}

The proposed in-cache computing paradigm is designed to address the challenges of high-latency memory access in data-intensive applications by enabling computational functionality directly within the \llc. 
In this context, \arcane offers two primary advantages:
\begin{enumerate*}
    \item it significantly reduces the impact of long latencies operation on the last memory level, effectively making them transparent to the system,
    \item it facilitates efficient execution of data-intensive tasks by leveraging cache resources, allowing \ooo program execution and improved overall system throughput. 
\end{enumerate*}

\begin{figure*}[ht]
\centering
\includegraphics[width=\linewidth]{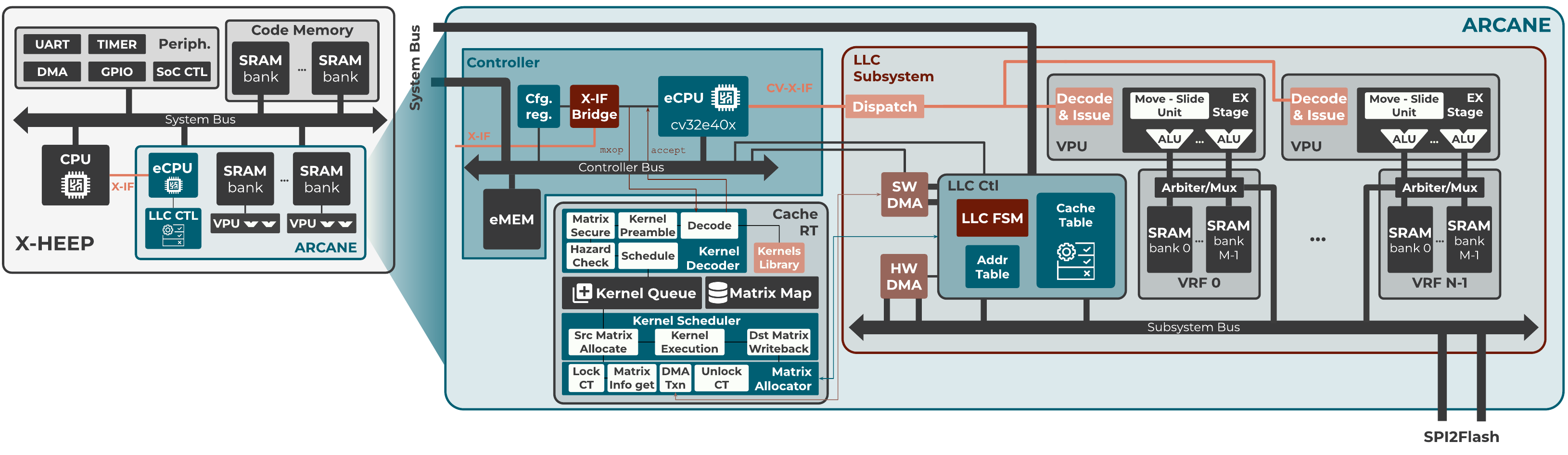}
\caption{X-HEEP system level block diagram with a detailed view of the \arcane \llc and software stack.}
\label{fig:block_diagram}
\end{figure*}

\arcane (\Cref{fig:block_diagram}) replaces a traditional data memory subsystem of a low-power \mcu with a smart \llc that operates both as a cache and computational unit. It is connected to the system bus with two slave ports, as well as ports towards external memories (e.g., flash or pseudo-static RAMs (PSTRAMs)).
In addition, the host \cpu can offload complex custom \riscv extensions for in-\llc operations that process cache data via the CV-X-IF \cite{cvxif}.
Thus, from the host CPU point of view, the proposed \llc operates both as a cache and as an external co-processor.
Computation within the \llc restricts the active cache region, introducing additional control complexity and potentially affecting performance.
However, in-cache computing can partially offset the delays caused by long-latency complex operations, thanks to (2).
Inside \arcane, to mitigate potential performance degradation, the \llc controller ensures efficient cache resources management, ensuring a balance between computational throughput and memory coherence.
Computation in the \arcane \llc is enabled by leveraging the \nmc paradigm, as presented in \cite{carus}. 
The custom in-\llc extensions are based on micro-programs built on top of the vector-like custom extensions proposed in the NMC IP called NM-Carus \cite{carus}.
Vector-like instructions are chosen to limit the overhead associated with control flow instructions and to utilize \ac{simd} capabilities to handle parallel workloads.
Differently from the work proposed in \cite{carus}, multiple NM-Carus instances managed by a single \ecpu are employed to build the \llc system. 
The \ecpu is based on the OpenHW Group's CV32E40X, a 4-stage, in-order \riscv core based on \cite{cv32e40p}, that offloads the custom vector-like near-memory operations to each NM-Carus instance, which act as \vpus.
A dispatcher carries out the distribution to the selected \vpus, keeping the architecture modular and scalable.
A slave port to the system bus facilitates the interaction between the host \cpu and the \emem to upload the \ecpu firmware and configure memory-mapped registers, ensuring flexibility in deployment and programmability.


The architecture of \arcane is analyzed in detail in \Cref{sec:llc}, focusing on its standard and in-cache computing functionalities. 
Furthermore, the offloading mechanism from the host \cpu to the \ecpu is discussed in \Cref{sec:bridge}, introducing the concept of software-decoded instructions, which further boosts the system's flexibility and usability.

\subsection{\arcane LLC}
\label{sec:llc}
\subsubsection{ Cache normal functioning mode}
The \llc is designed as a fully associative cache, with a total number of lines equal to the aggregate vector register capacity of the system (i.e., the number of \acp{vpu} by vector registers per \ac{vpu}).
The cache line length is configured to match the maximum supported vector size to streamline memory management and ensure coherence between computational and caching operations, avoiding memory fragmentation issues.
Cache hits are resolved in a single cycle, while misses and write-backs are handled by a dedicated \dma.
The \llc implements an approximate version of the \ac{lru} replacement policy using a counter-based approach to maintain effective cache utilization.
Furthermore, a write-back writing policy is enforced for improved performance, writing a cache line back to its original memory location on dirty replacement.
\subsubsection{Cache locking and hazards management}
In-cache computation enables parallel execution of the host \cpu and the \ecpu program flow. 
The cache controller must mediate between the \caos and the host \cpu to handle contention on the \vpus.
A locking mechanism is implemented through a memory-mapped configuration register, written by the \ecpu and read by the controller to synchronize cache accesses.
When the \ecpu acquires the lock, the host \cpu is blocked from accessing the cache until the lock is released. 
Conversely, a potential \ecpu's lock request is not granted during ongoing host \cpu operations, thus stalling the \caos until the memory operation concludes.

During kernel execution, cache regions allocated to kernel operands are marked with a \texttt{busy computing} status to prevent access by normal operations, thus ensuring cache consistency while enabling in-cache computing.
Potential hazards may arise from concurrent host \cpu and \ecpu operations.
\ac{war} hazards occur when the host \cpu issues a store operation on a source operand of an active kernel, potentially overwriting it before allocation is complete. As kernel operand allocation involves creating temporary copies in the \vpu cache lines arranged according to the kernel layout, a blocking mechanism must prevent store operations on the sources until allocation is finalized.
\ac{raw} hazards arise if the host \cpu reads the kernel result before the computation is complete, while \acp{waw} emerge if the kernel destination overwrites the result of a subsequent store. 
All operations targeting kernel destinations must be blocked until kernel write-back is completed to avoid conflicts.
The controller overcomes stalling conditions once memory contention resolves.
\subsubsection{\ac{at}}
The system employs this auxiliary table to manage kernel source and destination states, ensuring proper synchronization and preventing data corruption while maintaining high throughput for in-cache computation. 
Each \at entry contains the start and end addresses of the operands, along with a validity and a status flag.
The \ecpu updates the \at when matrices are registered and sets their status to \texttt{busy} according to the hazard-avoidance policy.
Additional status bits in the \ct indicate whether a cache line contains a source or a destination to streamline access.
This approach allows the \at to be checked only when a corresponding cache line is marked as a source or destination, keeping a one-cycle delay in case of a hit.
Cache misses always involve an \at lookup to determine if the request pertains to an allocated operand.
If the required element belongs to a critical cache line but is not part of an operand, the entire line is loaded from memory, marked as containing an operand data and the request is served. 
This mechanism preserves program correctness by stalling only critical memory requests while allowing non-critical operations to proceed.
\subsubsection{Software-Driven \ac{dma}}
During the kernel allocation phase, the \ecpu leverages X-HEEP's specialized \dma supporting 2D transactions, transferring operands from the main memory to the selected \vpu in the required matrix format. 
To optimize latency, the \llc controller updates cache entry statuses upon receiving a \dma request, bypassing the need for software to search the \ct.
The \dma requests, as depicted in \Cref{fig:block_diagram}, are routed through the \llc controller, which forwards data either from the cache (on a hit) or from the external off-chip memory (on a miss).
The controller updates hit lines associated with sources or destinations and marks the lines required for computing as \texttt{busy computing}, handling write-back if required.
During the kernel write-back, the cache follows a \textit{fetch-on-write} policy, updating destinations directly in the cache and marking the corresponding lines as \texttt{dirty}.

\subsection{Bridge}\label{sec:bridge}
%

%
%
%
The bridge shown in \Cref{fig:block_diagram} provides a unified interface between the host \cpu and \ecpu, enabling offloading of matrix operations via the \xif. Operations are processed by the \ecpu through an interrupt-driven, memory-mapped mechanism, completely transparent to the host \cpu. The bridge samples the instruction’s \texttt{opcode}, \texttt{func5}, and register operands coming from the offloaded \riscv instruction over the \xif, making data accessible to the \ecpu.
Upon offloading, the bridge raises an interrupt for the \ecpu to decode the instruction in SW. A dedicated \ecpu register logs the decoding outcome, which the bridge forwards to the host \cpu using \xif. The host \cpu then sends commit or kill signals through the bridge, which idles upon kill acknowledgment.
For operations proceeding to execution, the bridge notifies the host \cpu, allowing it to continue the application workflow in an \ooo fashion.

\section{Software-defined In-Cache Instruction Set Extensions}
\label{sec:sw}

Despite its advantages, the widespread adoption of NMC in data-driven applications (e.g., neural networks, signal processing) is often limited by the complexity programmers face when writing kernels. This process requires a deep understanding of the hardware architecture and the precise data layout within the memory hierarchy. These challenges can be mitigated by encapsulating a kernel into a single complex instruction, thereby offloading data management to the existing \cpu controller within the \nmc subsystem. The \cpu controller decodes the complex in-cache instructions in software, offering enhanced flexibility with negligible impact on application throughput,
as demonstrated in \Cref{sec:overhead}.
The kernel itself, which represents the micro-program that executes the complex in-cache instruction, is built by leveraging the custom near-memory vector-like \riscv extensions presented in \cite{carus}, which are instead decoded and executed in HW by the NMC instances.
This hierarchical approach makes programmability straight-and-forwards from the host \cpu point of view, which handles compact and well-defined complex in-cache instructions while offering at the same time flexibility thanks to the SW-defined instructions, and performance, thanks to the vector-like near-memory instructions implemented with parallel data-path.

The following section introduces a reconfigurable in-cache SW-defined \riscv matrix \isa that abstracts the complexity of NMC. It is followed by a detailed discussion of the software system residing within the NMC subsystem.

\begin{figure*}
    \begin{minipage}[hbt]{0.74\textwidth} 
        \vspace{0pt}
        \scriptsize
        \begin{tabular}{l|llllll|l}
            \toprule
            \multicolumn{1}{c|}{\textbf{Mnemonic}} &\multicolumn{6}{|c|}{\textbf{Data sources}} &\multicolumn{1}{|c}{\textbf{Description}}\\
            \multicolumn{1}{c|}{\textbf{}}
            &\multicolumn{1}{|c}{hi(rs1)} 
            &\multicolumn{1}{c}{lo(rs1)} 
            &\multicolumn{1}{c}{hi(rs2)} 
            &\multicolumn{1}{c}{lo(rs2)} 
            &\multicolumn{1}{c}{hi(rs3)} 
            &\multicolumn{1}{c|}{lo(rs3)} 
            &\multicolumn{1}{|c}{\textbf{}}\\
            \midrule
            \mintinline{gas}|xmr.[w, h, b]| & \mintinline{c}|hi(&A)|    & \mintinline{c}|lo(&A)| & \mintinline{gas}|A.stride|    & \mintinline{gas}|md| & \mintinline{gas}|A.cols| & \mintinline{gas}|A.rows|  &Matrix reserve\\
            \mintinline{gas}|xmk0.[w, h, b]|& $\alpha$    &$\beta$ & \mintinline{gas}|ms3|    & \mintinline{gas}|md| & \mintinline{gas}|ms1| & \mintinline{gas}|ms2|  &GeMM\\
            \mintinline{gas}|xmk1.[w, h, b]|&$\alpha$    & \mintinline{c}|-| & \mintinline{c}|-|    &\mintinline{gas}|md| &\mintinline{gas}|ms1| & \mintinline{c}|-|  & LeakyReLU\\
            \mintinline{gas}|xmk2.[w, h, b]|& \mintinline{gas}|stride|    & \mintinline{gas}|win_size| & \mintinline{c}|-|    & \mintinline{gas}|md| & \mintinline{gas}|ms1| & \mintinline{c}|-|  & Maxpooling\\
            \mintinline{gas}|xmk3.[w, h, b]|& \mintinline{c}|-|    & \mintinline{c}|-| &\mintinline{c}|-|    & \mintinline{gas}|md| & \mintinline{gas}|ms1| & \mintinline{gas}|ms2|   &2D Conv.\\
            \mintinline{gas}|xmk4.[w, h, b]|& \mintinline{c}|-|    & \mintinline{c}|-| &\mintinline{c}|-|    & \mintinline{gas}|md| & \mintinline{gas}|ms1| & \mintinline{gas}|ms2|  &3-ch. 2D Conv. Layer\\
            \bottomrule
        \end{tabular}
        \vspace{2.5mm}
        \captionof{table}{Example of \arcane custom kernels.}\label{tbl:isa}
    \end{minipage}%
    \hfill
    \begin{minipage}[hbt]{0.25\textwidth} 
        \centering
        \begin{minted}[
            label={Convolutional Layer},
            frame=lines,
            framesep=2mm,
            baselinestretch=1.0,
            bgcolor=lightgray,
            fontsize=\notsotiny,
            numbersep=5pt,
            xleftmargin=0pt
        ]{c}
// Convolutional Layer
int main(void) {
  int A[rowsA][colsA] = {...}
  ...
  // Reservation
  _xmr_w(m0, A, 1, rowsA, colsA);
  _xmr_w(m1, F, 1, rowsF, colsF);
  _xmr_w(m2, R, 1, rowsR, colsR);
  // Matrix Kernel
  _conv_layer_w(m2, m0, m1);
  ...
\end{minted}
        \captionof{listing}{\xisa application example.}\label{lst:kernel_examples}
    \end{minipage}
\end{figure*}

\subsection{Extendable In-Cache Matrix ISA}
The \xisa extension is implemented within the \riscv \emph{Custom-2} 25-bit
encoding space, utilizing the \texttt{0x5b} major opcode. To maximize the
utility of a single instruction, each source register is divided into 16-bit
pairs, with four registers allocated for matrix register indices and two
reserved for scalar parameters, α and β. This configuration accommodates
parameter-intensive kernels, such as General Matrix Multiplication (GeMM), while
maintaining flexibility. To ensure a high level of abstraction, the extension
includes only two types of instructions: 

\subsubsection{Matrix Reserve (\mintinline{gas}|xmr|)}
The \mintinline{gas}|xmr| instruction binds a matrix's memory address and shape to a logical
matrix register. Leveraging software-hardware cooperation within the LLC
subsystem, memory coherency is ensured with no additional effort from the
programmer. Unlike a matrix load instruction as proposed by the T-HEAD RVM proposal \cite{t_head}, \mintinline{gas}|xmr| does not immediately
load matrix data into memory. Instead, it establishes a binding, deferring the
memory load to a later stage, such as when it is explicitly requested by a
matrix kernel operation. This deferred approach abstracts memory management
complexities, significantly reducing the programmer's workload.

\subsubsection{Matrix Kernels (\texttt{xmkN})}
Matrix kernel instructions, denoted as xmkN where N$\in$[0,30], define up to 31
distinct complex matrix kernel operations. The func5 field within the
\texttt{0x5b} opcode specifies the kernel operation. This flexibility is further
enhanced by the reprogrammable software decoder, allowing for updates, and
further \ac{isa} extensions.

Internally, kernels represent matrices as groups of vector registers, maximizing
the reuse of the custom vector-like extension provided by the chosen NMC architecture \cite{carus}.
Currently, five complex matrix kernels have been implemented, some of which
inherited from existing implementations \cite{carus}.

To demonstrate the abstraction capabilities of the \xisa extension, a 3-channel
2D convolution kernel operation inspired by ImageNet has been implemented. This
kernel integrates 2D convolution, max-pooling, and ReLU activation while
supporting matrices of arbitrary dimensions. 

An example of using the \xisa is provided in \Cref{lst:kernel_examples}.

\lstdefinestyle{customc}{
    language=C,                              
    basicstyle=\ttfamily\tiny,             
    keywordstyle=\color[HTML]{6F1A07}\bfseries,     
    keywordstyle=[2]\color{red},   
    keywordstyle=[3]\color{black!60!black}, 
    keywordstyle=[4]\color[HTML]{287889}\bfseries, 
    identifierstyle=\color{blue},        
    commentstyle=\color{teal}\itshape,              
    frame=lines,                            
    morekeywords={xmr_w},             
    morekeywords=[2]{m0, m1, m2, m3, m4, m5, m6, m7}, 
    morekeywords=[3]{rowsA, rowsB, colsA, colsB, rowsC, rowsF, colsF, rowsR, colsR},  
    morekeywords=[4]{gemm_w, conv_layer_w},  
    captionpos=b, 
}

\subsection{Cache Runtime System}
\caos is a lightweight runtime system designed to
perform three core tasks: software decoding of matrix operations, their scheduling and
execution, and matrix allocation. It is executed only by the \ecpu within the \llc. These modules operate
independently but can communicate. \caos 
operates as a single-threaded, preemptive runtime, ensuring efficient handling of
offloaded matrix operations, even during kernel execution. It follows a
producer-consumer model centered around a statically allocated kernels queue.

\caos employs a static memory allocation philosophy, offering two key benefits:
predictable runtime without memory fragmentation and the ability to analyze
maximal stack usage. Critical structures, such as the kernel queue and the
matrix map, are preallocated to fixed sizes determined by system configuration.
For instance, the matrix map supports a configurable number of logical matrix
registers. A user-configurable kernel library allows custom kernels to be added
before \caos compilation. Additionally, \caos supports a deep-sleep mode for power
efficiency when no operations are pending.

The primary modules of \caos, are the Kernel Decoder, Kernel Scheduler, and Matrix
Allocator:

\subsubsection{Kernel Decoder}
Operates within the interrupt handler, decoding matrix
operations offloaded by the host \cpu. It retrieves kernel information, such as
preambles and function addresses, using the operation's \texttt{opcode} and \texttt{func5} field with $O(1)$ complexity access to the kernel library. If the operation is
recognized, the kernel preamble is executed, and upon success, the operation is
scheduled and added to the kernel queue. 
Due to ooo communication with the host \cpu, data-inbound
matrices may risk modification before subsequent computations. To address this,
the Kernel Decoder records the start and end addresses of the memory region in the \at,
preventing undesired reads/writes to a destination/source operand still required by pending matrix operations.
Additionally, to mitigate hazards such as having a \mintinline{gas}|xmr| overwriting an older
reservation still in use, the Kernel Decoder employs a hazard checker that
internally renames logical matrices effectively solving the hazard. Notably,
matrix allocation does not happen during \mintinline{gas}|xmr| execution. Instead, it is deferred
until explicitly required by a kernel operation, enabling kernel-dependent
layout optimization.

\subsubsection{Kernel Scheduler}
Manages the execution of matrix operations. Before
execution, it selects an appropriate \ac{vpu} based on a policy that prioritizes \acp{vpu} with the fewest dirty cache lines.
Once a \ac{vpu} is selected, the scheduler invokes the Matrix Allocator to prepare
source matrices with the required layout.
After kernel execution, the scheduler determines whether the destination matrix
will serve as a source operand in future operations. If not, it writes the
matrix back to memory using the Matrix Allocator \ac{api}.

\subsubsection{Matrix Allocator}
Handles memory management for matrix operations. It
receives the matrix operand layout information to program 2D \ac{dma} transfers that
move data from memory to the selected \ac{vpu}. To prevent clashes in cache line
access with the host \cpu, the allocator must first acquire a lock on the cache
controller. The Matrix Allocator further minimizes the overhead impact on throughput by allocating the effective dimensions of the matrix. A
custom \dma controller ensures data integrity by detecting writes on dirty cache
lines and triggering writebacks. After the allocation process is completed, the
Matrix Allocator releases the cache controller lock.
During the writeback phase, which occurs post-kernel execution, the Matrix
Allocator locks the \llc Controller and programs a 2D \dma transfer to consolidate
scattered matrix-shaped data into a contiguous array inside the \llc. If the cache line containing the matrix is not already present in the \llc, it is first loaded and then updated with the newly computed data from the matrix operation. This ensures that any pending access requests for the updated data can be promptly served with the latest data.
Once the transfer completes, it marks the previously busy cache lines as free
and releases the \llc Controller and memory region. The memory region is then
made accessible to the host \cpu by modifying the \at's access permissions.

\section{Experimental Results}
\label{sec:results}
\subsection{Logic Synthesis}
\begin{figure}[ht]
\centering
\includegraphics[width=\linewidth]{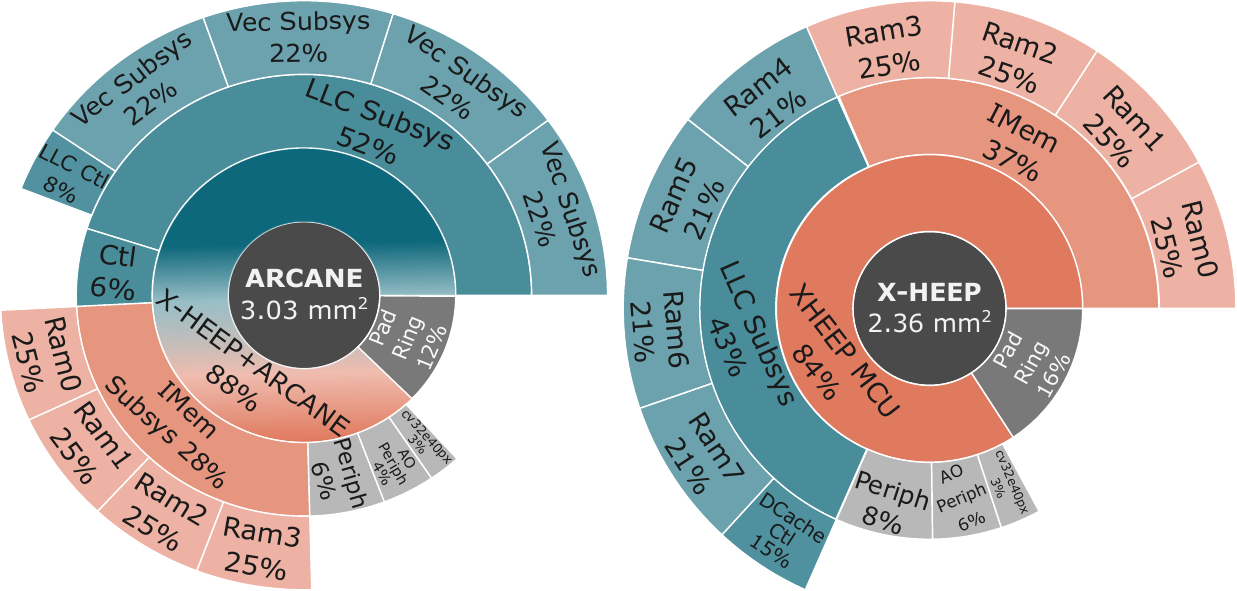}
\caption{Area split of \ac{xheep} + \arcane 4-lanes configuration (\qty{128}{\kilo iB}) versus \ac{xheep} + standard data \llc (\qty{128}{\kilo iB})}
\label{fig:arcane_vs_heep}
\end{figure}

The logic synthesis of \arcane is performed using Synopsys Design Compiler\textsuperscript{\textregistered} 2020.09, targeting a low-power \unit{65 \nano m} LP CMOS technology library under worst-case operating conditions with a target clock frequency of \qty{250}{\mega Hz}. 
\arcane is encapsulated within the \ac{xheep} \mcu framework replacing the conventional data memory. 
Three configurations, providing a range of design trade-offs to balance computational throughput and area overhead, are synthesized, differing in the number of lanes, and, consequently, the number of memory banks per \vpu.
Across all configurations, the \texttt{cv32e40x} core implementing the RV32IMC instruction set serves as the \ecpu, supported by a \qty{16}{\kilo iB} \emem. 
The \mcu instruction memory consists of 4 banks of \qty{32}{\kilo iB} each, for a total of  \qty{128}{\kilo iB}, while the data cache has a total capacity of \qty{128}{\kilo iB} and is split into 4 \vpus with a vector length and a cache line size of \qty{1}{\kilo iB}.
The \ac{xheep} system, featuring the \texttt{cv32e40px} core \cite{cv32e40p}, with identical instruction memory size and augmented with a standard data \llc, is used as a baseline for a fair comparison.
\Cref{tab:synth} compares the synthesized configurations, highlighting their area overhead with respect to the baseline.
The additional area introduced by \arcane primarily stems from its enhanced computational capabilities, as shown in \Cref{fig:arcane_vs_heep}, while the impact of the additional control logic for managing concurrent computation and memory operations remains negligible.
For the intermediate 4-lanes configuration, \arcane shows a 28.3\% area increase over the baseline, with 22\% attributed to the vector pipelines and 5\% to the controller, split among the \ecpu and the \emem.
As the number of lanes increases, area overhead grows due to more complex computation logic, reduced memory density from \llc division, and higher routing complexity.
Notably, the additional cache control logic accounts for less than 4\% of the total system area.

As a final remark, \arcane \llc does not increase the critical path of the target frequency of \qty{250}{\mega Hz}.
\begin{table}[h]
    \centering
        \caption[Synthesis results]{Synthesis results with \SI{16}{\kibi\byte} \ac{emem}.}
    \label{tab:synth}
    \resizebox{\columnwidth}{!}{
    \begin{threeparttable}
    \begin{tabular}{ccccc}
    \toprule
	 Conf & \multicolumn{1}{c}{\begin{tabular}[c]{@{}c@{}}\arcane\\ (4 \acp{vpu}\tnotemid{a}, 2 lanes)\end{tabular}} & \multicolumn{1}{c}{\begin{tabular}[c]{@{}c@{}}\arcane\\ (4 \acp{vpu}\tnotemid{a}, 4 lanes)\end{tabular}}  & \multicolumn{1}{c}{\begin{tabular}[c]{@{}c@{}}\arcane\\ (4 \acp{vpu}\tnotemid{a}, 8 lanes)\end{tabular}} & \multicolumn{1}{c}{\begin{tabular}[c]{@{}c@{}}X-HEEP\cite{heep}\\ (4 DMem Banks\tnotemid{a})\end{tabular}}  \\ \midrule[0.2pt]
     \multirow{2}{*}{Area\unit{[\micro m \squared]}}  & \num{2.88d6}  & \num{3.03d6} & \num{3.34d6}  & \num{2.36d6}\\
     & (+\num{21.7}\%) & (+\num{28.3}\%) & (+\num{41.3}\%) & \\ \midrule[0.2pt]
     Area\unit{[\kilo GE]\tnote{c}}  & 1996 & 2105  & 2318  & 1640 \\ \bottomrule
    \end{tabular}
    \small
    \begin{tablenotes}[]
        \item[a] \SI{32}{\kibi\byte} each.
        \item[c] \unit{GE} is the 2-input drive strength-one NAND gate equivalent area.
    \end{tablenotes}
    \end{threeparttable}
} 
\end{table}


\subsection{Overhead Analysis}
\label{sec:overhead}

\begin{figure}[htb]
\centering
\includegraphics[width=\linewidth]{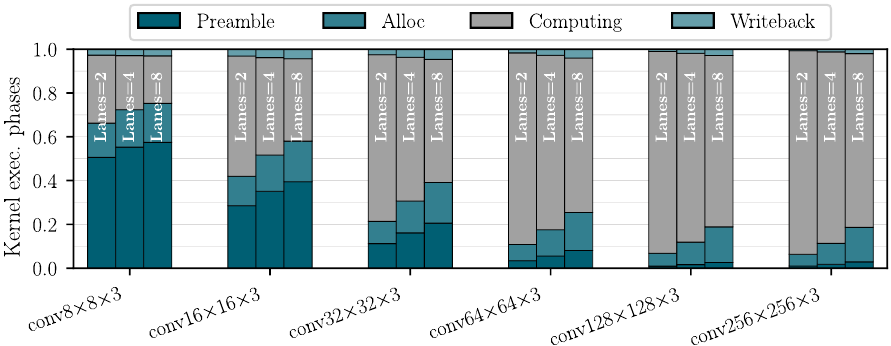}
\caption{Non-compute phases overhead analysis under different input matrix sizes and \arcane lanes with \texttt{int32} datatype.}
\label{fig:overhead}
\vspace{-5mm}
\end{figure}

Providing abstraction to programmers is a primary goal of the \xisa extension, achieved through software decoding, allocation, kernel execution, and writeback phases. In \cref{lst:kernel_examples}, multiple \mintinline{gas}|xmr| instructions define kernel operands in the preamble phase, deferring data loading to kernel execution. Data movement during allocation and writeback is handled by \dma transfers.
These phases introduce throughput overhead. A worst-case study of a 3-channel 2D convolution kernel with $3\times3$ filters on \texttt{int32} integers was conducted using 2-, 4-, and 8-lanes \arcane shown in \Cref{fig:overhead}. Preamble phase overhead decreases exponentially with input size, from 60\% for small inputs to 2.89\% for larger ones, making \arcane a suitable solution for relatively large input sizes. Allocation overhead grows with lane count, saturating globally at 15\%, proportional to the input size. Writeback overhead falls linearly with input size, reaching 2\% for the largest matrices.
As inputs increase the compute phase dominates as input size increases, with overhead saturating at 20\% under worst-case conditions.

\subsection{Comparison with the State of the Art}
\begin{figure*}[htb]
    \centering
    \includegraphics[width=\linewidth]{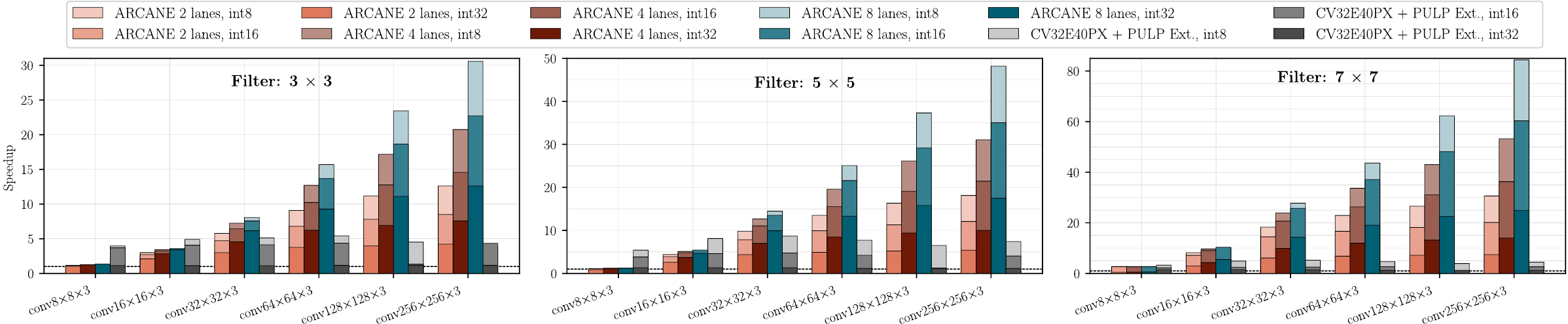}
    \caption{Speedup comparison between single instance \arcane configurations, CV32E40X and CV32E40PX featuring XCVPULP extensions.}
    \label{fig:speedup}
\end{figure*}

We compare \arcane with the state-of-the-art (\Cref{sec:related}), including a baseline CV32E40X \cpu core for speedup measurement and the CV32E40PX \cite{cv32e40p}, a core implementing the XCVPULP \isa extensions for 8- and 16-bit data, supported by the OpenHW Group toolchain.
We use a 3-channel 2D convolution layer (\Cref{lst:kernel_examples}), relevant to edge AI and tiny CNNs, for comparison with varying filter sizes, \arcane lane configurations, and data types. Results (\Cref{fig:speedup}) show that \arcane's 2-lane configuration achieves peak throughput at $64\times64$ inputs, saturating faster with larger filters due to lane limits.
In \texttt{int8}, prevalent in tinyML, performance saturation occurs in 2-lane setups, while 4- and 8-lane setups show consistent speedups with larger inputs due to optimized \dma transfers reducing allocation times. Although CV32E40PX outperforms \arcane at smaller input sizes, its scaling peaks at $8.6\times$ due to overhead from repeated data loading. \arcane's overhead is higher for smaller inputs but excels in processing large datasets. For instance, at $256\times256$ inputs with \texttt{int8} $3\times3$ filters, \arcane's 8-lane setup achieves a $30\times$ speedup over CV32E40X, compared to CV32E40PX's $5\times$.
In multi-instance mode with 4 VPUs and 8 lanes, \arcane achieves a $120\times$ speedup compared to CV32E40X and $1.6\times$ compared to CV32E40PX, with area utilization comparable to a 15-core CV32E40PX system, excluding logic and bus contributions. Multi-core implementations relying on packed-\acs{simd} instructions introduce significant overhead from frequent instruction cache accesses, causing memory contention and synchronization delays. Even under optimal conditions, the theoretical speedup peaks at $75\times$, far below \arcane's.

As motivated in \Cref{sec:related}, BLADE \cite{blade} and Intel CNC \cite{intel_cache} are selected as candidates for comparisons. Due to their restricted set of supported operations, running a direct comparison employing the 3-channel 2D convolution layer is not feasible. Nevertheless, it is possible to compare their peak throughputs. Given the difference in the technological nodes of such solutions, the results are scaled using an operational clock frequency of \SI{330}{\mega\hertz}, typical value of an embedded \SI{32}{\kibi\byte} SRAM in the \SI{65}{\nano\meter} node. 
BLADE \cite{blade} implements a \nmc architecture with a scaled area of $580\times10^3$ \SI{}{ \micro\meter}$^2$, making it $3.18\times$ smaller than \arcane. However, its peak throughput is limited to 5.3 \gops
\footnote{One MAC operation is considered as two OP (one multiplication and one addition), as commonly done in literature.}, 
while \arcane, running at \SI{265}{\mega\hertz}, achieves a peak throughput of 17.0 \gops—an improvement of approximately $3.2\times$. This translates to an area efficiency of 9.1 \gops/\SI{}{\milli\meter}$^2$ for BLADE, compared to 9.2 \gops/\SI{}{\milli\meter}$^2$ for \arcane, which demonstrates slightly superior area efficiency and also supports an extensible \isa while BLADE’s is restricted to basic arithmetic operations.
Intel CNC \cite{intel_cache}, fabricated using Intel’s 4 technology node, renders scaled area comparisons impractical due to substantially different fabrication technologies. Nevertheless, its reported area is $1920\times10^3$ \SI{}{\micro\meter}$^2$, larger than that of \arcane in \SI{65}{\nano\meter} node. The peak throughput of Intel CNC is 25.0 \gops, achieving a $1.47\times$ speedup compared to \arcane. Despite this, \arcane still demonstrates greater programming flexibility, as Intel CNC supports only the MAC operation.

\section{Conclusion \& Future Works}
\label{sec:conclusion}
The \arcane in-cache computing architecture combines the energy and performance benefits of the \nmc paradigm with the programmability of a \cpu-based solution. It does so by leveraging a \riscv cache controller and a customizable, software-defined \isa that operates as an abstraction layer to offload complex matrix operations to the cache. \arcane is a drop-in replacement for a system's \ac{llc}, doubling as a programmable matrix coprocessor. The cache controller relieves the application software from explicitly managing data movement and synchronization, enabling straightforward software development. When executing an 8-bit $256 \times 256 \times 3$ convolutional layer with a $7 \times 7$ filter, \arcane achieves a performance improvement of \SI{84}{\times} over a scalar \cpu implementation (RV32IIMC) and \SI{16}{\times} over the XCVPULP packed-\ac{simd} and DSP-enhanced \isa. It achieves comparable performance to existing solutions with a low area overhead of \SI{41.3}{\percent} when integrated into an edge-oriented \ac{mcu}.

\begin{acks}
This work is supported by the EU TRISTAN project with GA 101095947, which has received funding from the CHIPS Joint Undertaking and its members, and including top-up funding by \textit{Ministero dello sviluppo economico}, and SERICS (PE00000014), under the MUR National Recovery and Resilience Plan funded by the European Union - NextGenerationEU.
\end{acks}


\begin{acronym}[XXXXX]\itemsep0pt
    \acro{alu}[ALU]{Arithmetic Logic Unit}
    \acro{ann}[ANN]{Artificial Neural Network}
    \acro{api}[API]{Application Programming Interface}
    \acro{asic}[ASIC]{Application-Specific Integrated Circuit}
    \acro{at}[AT]{Address Table}
    \acro{bpu}[BPU]{Branch Prediction Unit}
    \acro{btb}[BTB]{Branch Target Buffer}
    \acro{bu}[BU]{Branch Unit}
    \acro{ct}[CT]{Cache Table}
    \acro{cdb}[CDB]{Common Data Bus}
    \acro{cim}[CIM]{Compute-In-Memory}
    \acro{clc}[CLC]{Configurabe-Latency Coprocessor}
    \acro{cos}[COS]{Cache Operating System}  
    \acro{cpu}[CPU]{Central Processing Unit}
    \acro{csr}[CSR]{Control and Status Register}
    \acro{dc}[DC]{Design Compiler}
    \acro{dlp}[DLP]{Data-Level Parallelism}
    \acro{dma}[DMA]{Direct Memory Access}
    \acro{dpu}[DPU]{Dot-Product Unit}
    \acro{eu}[EU]{Execution Unit}
    \acro{fll}[FLL]{Frequency-Locked Loop}
    \acro{fp}[FP]{Floating Point}
    \acro{fpga}[FPGA]{Field-Programmable Gate Array}
    \acro{fpu}[FPU]{Floating Point Unit}
    \acro{gpr}[GPR]{General Purpose Register}
    \acro{hdl}[HDL]{Hardware Description Language}
    \acro{ilp}[ILP]{Instruction-Level Parallelism}
    \acro{imc}[IMC]{In-Memory Computing}
    \acro{iot}[IoT]{Internet of Things}
    \acro{ipc}[IPC]{Instructions Per Cycle}
    \acro{iq}[IQ]{Issue Queue}
    \acro{isa}[ISA]{Instruction Set Architecture}
    \acro{lp}[LP]{Low Power}
    \acro{mac}[MAC]{Multiply-and-Accumulate}
    \acro{mcu}[MCU]{Microcontroller Unit}
    \acro{nib}[NIB]{Number of Instructions per Block}
    \acro{nmc}[NMC]{Near-Memory Computing}
    \acro{nn}[NN]{Nural Network}
    \acro{ooo}[OoO]{Out-of-Order}
    \acro{os}[OS]{Operating System}
    \acro{pc}[PC]{Program Counter}
    \acro{pqc}[PQC]{Post-Quantum Cryptography}
    \acro{lb}[LB]{Load Buffer}
    \acro{llc}[LLC]{Last Level Cache}
    \acro{lru}[LRU]{Least Recently Used}
    \acro{lsu}[LSU]{Load-Store Unit}
    \acro{noc}[NoC]{Network on Chip}
    \acro{ras}[RAS]{Return Address Stack}
    \acro{raw}[RAW]{Read-After-Write}
    \acro{rf}[RF]{Register File}
    \acro{rob}[ROB]{ReOrder Buffer}
    \acro{rom}[ROM]{Read-Only Memory}
    \acro{rs}[RS]{Reservation Station}
    \acro{rvv}[RVV]{RISC-V Vector Extension}
    \acro{rtl}[RTL]{Register Transfer Level}
    \acro{sb}[SB]{Store Buffer}
    \acro{simd}[SIMD]{Single Instruction Multiple Data}
    \acro{soc}[SoC]{System-on-Chip}
    \acroplural{soc}[SoCs]{Systems on Chip}
    \acro{sram}[SRAM]{Static Random-Access Memory}
    \acro{vpu}[VPU]{Vector Processing Unit}
    \acro{war}[WAR]{Write-After-Read}
    \acro{waw}[WAW]{Write-After-Write}
    \acro{x-heep}[X-HEEP]{eXtendable Heterogeneous Energy-efficient Platform}
    \acro{ecpu}[eCPU]{embedded CPU}
    \acro{emem}[eMEM]{embedded Memory}
    \acro{xif}[CV-X-IF]{CORE-V-X-IF}
    \acro{xheep}[X-HEEP]{eXtendible Heterogeneous Energy-Efficient Platform} 
    \acro{cos}[C-RT]{Cache Runtime}
    \acro{gops}[GOPS]{GOPS} 
\end{acronym}


\bibliographystyle{IEEEtran}
\bibliography{references}

\end{document}